\documentclass[prl,groupedaddress,twocolumn]{revtex4}
\usepackage[dvips]{graphicx}

\begin{document}

\bibliographystyle{apsrev}

\title{Energy relaxation during hot--exciton transport in quantum wells: Direct observation by spatially resolved phonon--sideband spectroscopy}
\author{Hui Zhao}
\author{Sebastian Moehl}
\author{Heinz Kalt}

\affiliation{Institut f\"{u}r Angewandte Physik, Universit\"{a}t Karlsruhe, D-76128 Karlsruhe, Germany}

\begin{abstract}

We investigate the energy relaxation of excitons during the
real--space transport in ZnSe quantum wells by using
microphotoluminescence with spatial resolution enhanced by a solid
immersion lens. The spatial evolution of the LO--phonon sideband,
originating from the LO--phonon assisted recombination of hot
excitons, is measured directly. By calculating the LO--phonon
assisted recombination probability, we obtain the nonthermal
energy distribution of excitons and observe directly the energy
relaxation of hot excitons during their transport. We find the
excitons remain hot during their transport on a length scale of
several micrometers. Thus, the excitonic transport on this scale
cannot be described by classical diffusion.

\end{abstract}

\maketitle

Excitons and their dynamics determine most of the linear optical
properties of a semiconductor quantum well (QW). For nonresonant
excitation, excitons are generated in a QW with a nonzero
center--of--mass kinetic energy. Such a hot--exciton injection is
followed by the evolution of the hot--exciton distribution in
energy space and real space towards equilibrium, i.e., by
relaxation and transport. The excitonic transport can be divided
into three essentially different regimes according to the relevant
status of the relaxation, namely the transport before, during, and
after the relaxation. The first regime extends up to the first
step of the relaxation, i.e., the first phonon--emission event.
Till then the exciton energy is conserved and one talks about
ballistic transport (when no scattering occurs at all) or
quasiballistic transport (only elastic scattering
occurs).\cite{bookmitin} The duration of this process is as short
as several picoseconds, determined by the phonon scattering time.
The last transport regime starts when the excitons have reached
quasi--equilibrium with the lattice. Since the typical relaxation
time of hot excitons is several 100~ps,\cite{b571390,jcg184795}
this transport regime can be observed only if the lifetime of the
excitons is long enough. Because the energy distribution remains
constant, one can apply a classical description using the
diffusion equation.\cite{b4715601}. Up to now, this classical
diffusion has been assumed to be the dominant process in excitonic
transport in
QWs.\cite{apl531937,b3910901,b423220,b451240,jap81536,apl74741,jap843611,apl74850}
In this letter, we will show that generally the second regime,
i.e., the transport during relaxation, dominates the excitonic
transport. Such a process is composed of the simultaneous
evolution of the exciton distributions in energy space and real
space.

Hot excitons are difficult to be studied by conventional optical
techniques. Since the photon momentum is very small, the direct
coupling of the hot exciton with a much larger momentum to the
photon is forbidden by momentum conservation. Only cold excitons
that have finished the relaxation can couple to photons, resulting
in the zero--phonon line (ZPL) in the photoluminescence (PL)
spectrum. This invisibility of the hot excitons in the spectrum is
the main obstacle of hot--exciton studies in conventional PL
experiments, which exploit the ZPL.

In this letter, we show that an LO--phonon sideband (PSB) can be
used as an ideal tool for direct investigation of the hot
excitons. By performing spatially resolved PSB spectroscopy, we
observe directly the energy relaxation of hot excitons during
their real--space transport in ZnSe QWs. We find that the excitons
remain hot on a length scale of several micrometers and that their
energy distribution remains nonthermal during their
lifetime--limited transport. Our result shows clearly that for
nonresonant excitation, the excitonic transport cannot be
described by classical diffusion.

The sample is a ZnSe~(7.3~nm)/ZnSSe~(10.7~nm) multiple QW grown by
Metal organic vapor phase epitaxy. Figure 1 shows schematically
the exciton formation and relaxation processes of the
QW.\cite{b571390} After an optical excitation in the continuum,
the generated electron--hole pairs rapidly form excitons assisted
by LO--phonon emission within few
picoseconds.\cite{l67128,b439354} The hot excitons then relax to
the band minimum by acoustic--phonon emission. After relaxation,
the cold excitons recombine radiatively, resulting in the ZPL.
From ZPL spectroscopy, we have obtained indirect hints for the
hot--exciton behavior.\cite{apl801391} As we discussed earlier,
the hot excitons cannot directly couple to the photon due to the
momentum difference. However, the LO phonon
\begin{figure}
 \includegraphics[width=7cm]{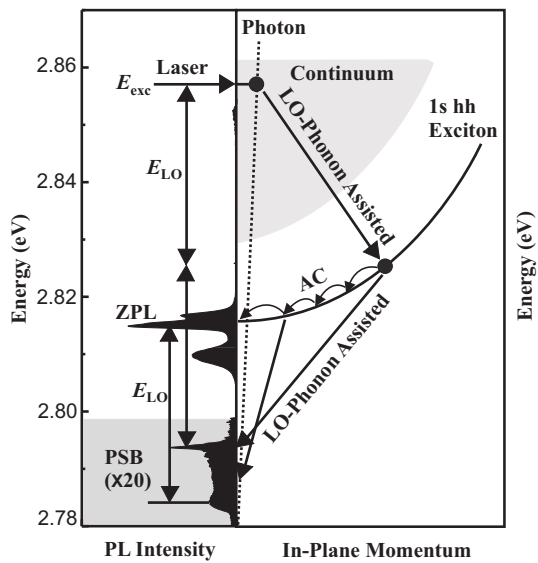}
 \caption{Excitonic process and photoluminescence of ZnSe quantum
 well. Left part:~A photoluminescence spectrum measured at a
 sample temperature of 7~K. The spectrum is composed of a
 zero--phonon line (ZPL) and a LO--phonon sideband (PSB);
 Right part:~Schematic drawing of the
 exciton formation assisted by LO--phonon emission and the subsequent
 relaxation by acoustic--phonon (AC) emission.}
\end{figure}
can assist the hot exciton in coupling to the photon by taking
away its momentum. The simultaneous well--defined energy loss to
the LO phonon leads to the appearance of the PSB, as shown in
Fig.~1. Since the spectral shape of the PSB reflects the
kinetic--energy distribution of
excitons,\cite{b571390,jcg184795,bookexciton} this LO--phonon
assisted recombination process provides an ideal tool for direct
investigation of hot excitons. Figure~2 shows several spectra of
the PSB excited by a cw laser at various photon energies
$E_{\mathrm{exc}}$. The PSB always appears in the photon--energy
range between $1E_{\mathrm{LO}}$ (LO--phonon energy, 31.8~meV in
this sample) below the ZPL and $2E_{\mathrm{LO}}$ below the
$E_{\mathrm{exc}}$.
\begin{figure}
 \includegraphics[width=6.5cm]{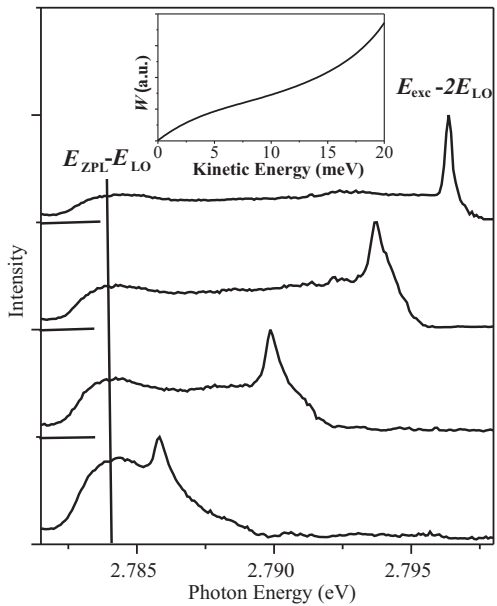}
 \caption{
  The spectra of the PSB excited by a continuous--wave laser at a sample temperature of 7~K.
  The excitation laser photon energies are (from top to bottom) 2.8602, 2.8573, 2.8535 and 2.8494~eV,
  respectively. Inset: The calculated LO--phonon assisted recombination probability $W$
  as a function of the exciton kinetic energy.
 }
\end{figure}

The shape of the PSB is related to the energy distribution of
excitons, $f(\varepsilon)$, by the LO--phonon assisted
recombination probability, $W(\varepsilon)$, i.e.,
$f(\varepsilon)=\mathrm{PSB}(\varepsilon)/W(\varepsilon)$. In
order to get the energy distribution from the PSB, we calculate
$W(\varepsilon)$. In the case of first order LO--phonon assisted
luminescence, the recombination probability is\cite{bookexciton}
\begin{equation}
W(\varepsilon)\propto\arrowvert\sum_{i}\frac{\psi_{i}(0)
H}{\varepsilon-E_{\mathrm{LO}}}\arrowvert^{2},
\end{equation}
where $\arrowvert\psi_{i}(0)\arrowvert^{2}$ is the oscillator
strength of the optical transition from the state $i$ in dipole
approximation. The summation is over all possible intermediate
states. For $1s$ excitons, the main intermediate state is $1s (K
\approx 0)$. $H$ is the Fr\"ohlich scattering matrix element.
Since we are only interested in the energy dependence of the
recombination probability, we omit the energy--independent
parameters, and get
\begin{equation}
W(\varepsilon)\propto\arrowvert\frac{[1+(a_{0}q_{c}/2)^{2}]^{-2}-[1+(a_{0}q_{v}/2)^{2}]^{-2}}{q(\varepsilon-E_{\mathrm{LO}})}\arrowvert^{2}.
\end{equation}
Here $a_{0}$ is the exciton Bohr radius, $q$ is the wave vector of
the LO phonon, $q_{c,v}=q(\mu/m_{c,v})$, where
$\mu=m_{c}m_{v}/(m_{c}+m_{v})$. The inset of Fig.~2 shows the
calculated result by using the parameters of ZnSe QW. We find an
increase of the recombination probability with the exciton kinetic
energy.

The spatially resolved PSB spectroscopy is performed with a solid
immersion lens (SIL)--enhanced confocal $\mu$--PL system. The
details of the system has been reported
previously.\cite{apl801391} The sample is excited locally by a
tunable cw laser through the objective, with an excitation spot of
about 200~nm half width at half maximum. The size of the
pinhole--defined detection spot is 460~nm in diameter. Although
the resolution of this system is lower than that of a scanning
near field optical microscope (SNOM) using coated fiber tips, the
collection efficiency is much higher than for the latter. Since
the PSB is much weaker than the ZPL (see Fig.~1), a high
collection efficiency is paramount for the PSB spectroscopy. One
can improve the collection efficiency of a SNOM by using an
uncoated fiber tip,\cite{apl76203} but simultaneously the spatial
resolution drops. Furthermore, a SNOM experiment has another
disadvantage: one cannot detect spectra from positions outside of
the excitation spot since local excitation and detection are
achieved by the same fiber tip. In contrast, by moving the pinhole
in the image plane of the microscope we can detect spectra from
positions outside of the excitation spot in a well--defined way.
This enables us to investigate the transport rather directly.

By scanning the pinhole thus varying the distance between the
excitation and detection spots, $d$, we detect the spatially
resolved PSB spectra, as shown in the left part of Fig.~3. With
increasing the $d$, we observe the change of the PSB spectral
shape. From these PSB spectra, we deduce the kinetic--energy
distributions of the excitons as shown in the right part of
Fig.~3.
\begin{figure}
 \includegraphics[width=6.5cm]{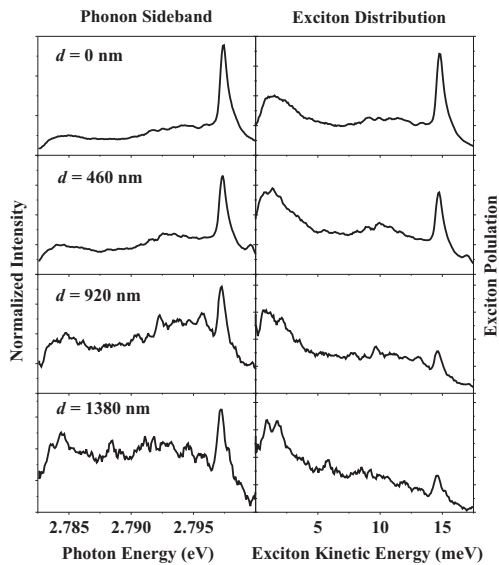}
 \caption{Left part: The spectra of PSB measured at different positions $d$ with respect to
 the excitation laser spot. The excitation intensity is 1~$\mathrm{kW/cm^{2}}$. The sample
 temperature is 7~K; Right part: The corresponding kinetic--energy distributions of excitons deduced from the spectra of PSB.
 }
\end{figure}
The energy relaxation of the nonthermal excitons during the
transport is clearly observed. We note that after a transport of
1380~nm, the exciton distribution is still nonthermal.

From the kinetic--energy distribution, we calculate the average
kinetic energy of the excitons. The spatial profiles of the
average kinetic energy are shown in Fig.~4
\begin{figure}
 \includegraphics[width=6.5cm]{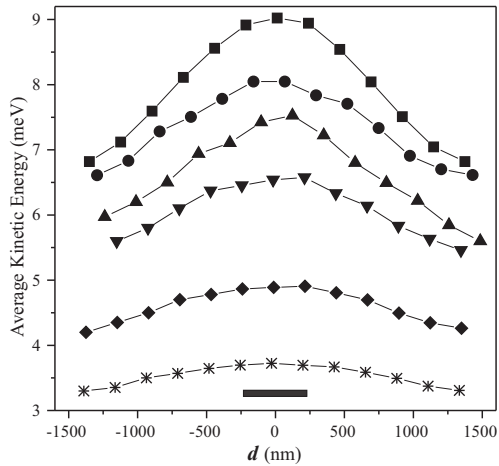}
 \caption{The average kinetic energy of excitons as function of
 $d$, the distance between the excitation and the detection spots. The initial kinetic energies of the excitons
 are (from top to bottom) 20.7, 17.6, 14.3, 11.0, 7.9 and 4.8~meV, respectively. The bar
 above the horizontal axis indicates the FWHM of the excitation laser spot.
 }
\end{figure}
for several values of $E_{\mathrm{ini}}$, the initial kinetic
energy of the excitons for the relaxation and transport processes.
Such a quantity is well--defined by $E_{\mathrm{exc}}$ (see
Fig.~1). We find the reduction of the average energy is only
$\sim20~\%$ after a transport of about 1500~nm. By comparing the
slopes of the curves in Fig.~4, we also find that the spatial
energy relaxation is slower for excitons with less kinetic energy.
This feature is consistent with the fact that the rate of the
acoustic--phonon emission, being the dominant relaxation
mechanism, increases with the exciton energy.\cite{b316552} In our
experiment, we cannot measure the average energy of excitons at a
position further than 1500~nm from the excitation spot since the
signal is too weak. However, we can conclude from the earlier two
facts that the excitonic transport on a length scale of at least
several micrometers is coupled to the energy relaxation. Although
the experiment is done on ZnSe QWs, the conclusion is
qualitatively general among other structures. In GaAs QW, the
hot--exciton transport feature is anticipated to be even more
pronounced due to the weaker Fr\"ohlich coupling (thus longer
energy--relaxation time) and the generally higher sample quality
(e.g., better interface quality thus higher mobility). Up to now,
excitonic transport in QWs after hot--exciton injection was
generally discussed in terms of classical diffusion. For example,
excitonic transport experiments of GaAs QWs performed on
micrometer length scale were modelled by diffusion equation,
although the carriers are excited with an excess energy of several
100~meV.\cite{apl531937,b3910901,b423220,b451240} In
investigations of ZnSe based QWs, diffusion length and diffusivity
of excitons were also deduced based on diffusion
equations.\cite{jap81536,apl74741,jap843611} In the InGaN/GaN
system, the transport within 1~$\mu$m was analyzed by neglecting
relaxation/thermalization processes.\cite{apl74850} Our experiment
shows clearly that in the case of nonresonant excitation, the
excitons remain hot and nonthermal on a length scale of several
micrometers, thus the transport is not a classical diffusion.

In summary, we show that the LO--phonon assisted recombination
process can be used for direct investigation of hot excitons. By
performing spatially resolved PSB spectroscopy, we observe
directly the spatial evolution of the nonthermal kinetic--energy
distribution of excitons. We find that at low temperatures the
excitonic transport on a length scale of several micrometers is
coupled to the energy relaxation, thus cannot be described as a
classical diffusion process.

We gratefully acknowledge the growth of excellent samples by the
group of M. Heuken (RWTH Aachen), and useful discussion with
H.~Giessen (Universit\"at Bonn). This work was supported by the
Deutsche Forschungsgemeinschaft (DFG) and the Center of Functional
Nanostructures (CFN).

\end{document}